\documentclass[12pt]{article}
\usepackage{graphicx,amsmath}
\usepackage{units}
\usepackage[usenames]{color}

\newlength{\dinwidth}
\newlength{\dinmargin}
\renewcommand{\vec}[1]{\boldsymbol{#1}}

\def\lapproxeq{\lower .7ex\hbox{$\;\stackrel{\textstyle                                                    
<}{\sim}\;$}}                                                    
\def\gapproxeq{\lower .7ex\hbox{$\;\stackrel{\textstyle                                                    
>}{\sim}\;$}}                                                    
\def\be{\begin{equation}}                                                    
\def\ee{\end{equation}}                                                    
\def\bea{\begin{eqnarray}}                                                    
\def\eea{\end{eqnarray}}
\def\b{\vec{b}}
     
\def\q{\vec{q}}

\def\GeV{\rm GeV}

\def\sh{\hat s}
\def\sh2{{\hat s}^2}

\def\el{\sigma_{\rm el}}
\def\DD{\sigma_{\rm DD}}
\def\SD{\sigma_{\rm SD}}

\begin{document}

\titlepage                                                    
\begin{flushright}                                                    
IPPP/13/81  \\
DCPT/13/162 \textbf{}\\                                                    
\today \\                                                    
\end{flushright} 
\vspace*{0.5cm}
\begin{center}                                                    
{\Large \bf High Energy Elastic and Diffractive Cross Sections}\\

\vspace*{1cm}
 V.A. Khoze$^{a,b}$, A.D. Martin$^a$ and M.G. Ryskin$^{a,b}$\\                                                 
\vspace*{0.5cm}                                                    
$^a$ Institute for Particle Physics Phenomenology, University of Durham, Durham, DH1 3LE \\                                                   
$^b$ Petersburg Nuclear Physics Institute, NRC Kurchatov Institute, Gatchina, St.~Petersburg, 188300, Russia

\vspace*{1cm}    
\begin{abstract}

We present a `global' description of the wide variety of high energy elastic and diffractive data that are presently available, particularly from the LHC experiments. The model is based on only one pomeron pole, but includes multi-pomeron interactions and, {\it significantly}, includes the transverse momentum dependence of intermediate partons as a function of their rapidity, which provides the rapidity dependence of the multi-pomeron vertices.
We give predictions for diffractive observables at LHC energies.

\end{abstract}                                                        
\vspace*{0.5cm}                                                    
                                                    
\end{center}

\section{Introduction}
 High energy diffractive processes caused by pomeron exchange are usually described 
within the framework of Reggeon Field Theory (RFT)~\cite{Gribov}. In the simplest case the high energy elastic scattering amplitude (and correspondingly the total cross section) is parametrised by single pomeron exchange, where the trajectory of this effective 
`soft' pomeron reads\footnote{More recent fits slightly change the values of $\Delta$ and $\alpha'_P$, but the precise values are not important for our discussion.}
\be
\alpha_P(t)=1+\Delta +\alpha'_P t\ ,
\label{eq:DL}
\ee 
with $\Delta=0.08$ and $\alpha'_P=0.25$ GeV$^{-2}$~\cite{DL}.

However, already two-particle $s$-channel unitarity generates a series of  
non-enhanced multi-pomeron diagrams leading to the eikonal approximation, in which the elastic amplitude in impact parameter, $b$, space is of the form
\be
\label{eq:U1}
T_{\rm el}~=~i(1-e^{-\Omega/2})\ ,
\ee
where the opacity $\Omega(s,b)$ plays the role of the phase-shift, $\delta_l$, of the (partial wave) amplitude with orbital momentum $l=b\sqrt s/2$ with $\Omega/2=2i\delta_l$. To be precise, (\ref{eq:U1}) is the solution of the $s$-channel unitarity equation.
\be
2{\rm Im}\ T_{\rm el}(b)= |T_{\rm el}(b)|^2+G_{\rm inel}(b)\ ,
\ee
where $G_{\rm inel}$ is the sum of the inelastic contributions.
 As usual, $\sqrt{s}$ is the c.m. energy. Hence we have
\bea \sigma_{\rm tot}(s,b) & = & 2(1-{\rm e}^{-\Omega/2})\\
\sigma_{\rm el}(s,b) & = & (1-{\rm e}^{-\Omega/2})^2, \label{eq:el}\\
\sigma_{\rm inel}(s,b) & = & 1- {\rm e}^{-\Omega}. 
\label{eq:inel}
\eea

To account for the possibility that the proton may  dissociate into low mass states, such as $p\to N^*$, we follow Good-Walker~\cite{GW} and introduce a multi-channel eikonal. That is, we decompose the proton state, $|p\rangle$ into the G-W diffractive eigenstates $|\phi_i\rangle$ ($|p\rangle=\sum a_i|\phi_i\rangle$) which undergo elastic scattering only,
 \be \langle \phi_i|T|\phi_k\rangle = 0\qquad{\rm for}\ i\neq k \ , 
\ee
leading to a multi-channel eikonal $\Omega_{ik}$, where the indices $i$ and $k$ now correspond to the beam and to the target protons.

Such an approach accounts for the rescattering of the incoming partons. However, from the microscopic point of view, pomeron exchange is described by a set of ladder-type diagrams. (This is true for both the `hard' BFKL pomeron and the `soft' multiperipheral pomeron.)  So we cannot exclude the rescattering of the intermediate
partons (produced during the evolution inside this ladder). In terms of  RFT these effects are described by triple- and  multi-pomeron vertices (with couplings $g^1_2$ and $g^n_m$ respectively); that is by the pomeron-pomeron interactions.

Recall that in conventional RFT it was assumed that all the transverse momenta are limited and that the Reggeon trajectories and the  couplings (including those for the multi-pomeron vertices)  do not depend on incoming energy, $\sqrt{s}$. This framework allowed a satisfactory description of the available diffractive data up to the Tevatron energy. However the new LHC data start to signal some problems\footnote{Some of these problems are also noted in Ref. \cite{GLM}.}:
\begin{itemize}
\item The total cross section and the $t$-slope of elastic scattering grow a bit faster than was expected based on the simplified DL parametrisation (\ref{eq:DL}). 
Indeed, the DL fit predicts $\sigma_{\rm tot}=90.7$ mb at $\sqrt{s}=7$ TeV, while TOTEM
observes 98.6$\pm 2.2$ mb \cite{TO1a}. The elastic slope was measured
at the Tevatron ($\sqrt s=1.8$ TeV) to be $B_{\rm el}=16.3\pm 0.3$ GeV$^{-2}$ by
the E710  experiment \cite{EEE} and to be
$B_{\rm el}=16.98\pm 0.25$ GeV$^{-2}$ by the CDF group \cite{cdfB}. Even starting from
the CDF result, and using the $\alpha'_P=0.25$ GeV$^{-2}$, we expect
 $B_{\rm el}=16.98+4\times 0.25\times \ln(7/1.8)=18.34$ GeV$^{-2}$ at 7 TeV, while
TOTEM finds $19.9\pm 0.3$ GeV$^{-2}$ \cite{TO1a}.

\item On the other hand, the preliminary values of cross section of diffractive dissociation, measured by TOTEM, turns out to be lower than that expected based on conventional RFT.

\item  Simultaneously, a growth of the mean transverse momenta of secondaries, with collider energy, is observed.
\end{itemize}

In the present paper we study, within RFT, the possibility that the transverse momentum, $k_t$, increases with energy. Can the growth of $k_t$ explain
the new features of the diffractive events observed at the LHC? The aim is not to reach a perfect quantitative description of the experimental data, but rather to understand the characteristic properties of high energy strong interactions.

Bearing in mind the relatively small values of the triple- and multi-pomeron couplings, we start with the simplest Reggeon diagrams. We include the absorptive (gap survival) effects caused by the eikonal and we consider the role of the increasing transverse momenta, which leads to a decrease of the
pomeron (and multi-pomeron) couplings, which are proportional to $ 1/k_t$. To make the discussion more transparent, we will not include explicitly the enhanced diagrams (which account for the rescattering of the intermediate ladder partons). The role of these
diagrams is mainly to renormalize (diminish) the intercept of the original (bare) pomeron and to enlarge the characteristic transverse momentum which arises from the stronger absorption of the partons with low $k_t$. Therefore we will use renormalised parameters of the pomeron trajectory (determined by fitting to the data),
 and a reasonable assumption for the energy and rapidity behaviour
of $k_t$.

\section{The high energy diffractive data}

At the moment, data for diffractive processes are available at 7 TeV, mainly from the TOTEM collaboration. TOTEM have measured the total and elastic cross sections (in a wide $t$ interval including the dip region)~\cite{TO1,TO1a}, the cross section of a low-mass ($M_X <3.4$ GeV) diffractive single ($pp\to p+X$)~\cite{TO2} and double ($pp\to X_1+X_2$)~\cite{TO3} dissociation; and made preliminary measurements of high-mass single proton dissociation, $\sigma_{\rm SD}$, integrated over the three intervals of $M_X$: namely $(3.4,8);~(8,350);~(350,1100)$ GeV~\cite{TO4}. In addition we have the inelastic cross sections and the cross sections of events with a Large Rapidity Gap (LRG) measured by the ATLAS \cite{atl}, CMS \cite{CMSdiff} and ALICE \cite{ALICE} collaborations. 

Formally the data from different groups do not contradict each other, since they are measured for different conditions. However there appear to be several tensions between the data sets.. 
\begin{itemize}
\item First, it is not easy to accommodate simultaneously the TOTEM result for $\sigma_{\rm SD}$ and the yield of LRG events observed by  ATLAS/CMS; see the discussion  in Sections 4.2 and 5.3 and in footnote 12. An analogous problem is  
described in the next bullet point below.

\item Moreover, the TOTEM $\sigma_{\rm SD}$ cross section looks too small in comparison with the value of $d\sigma_{\rm SD}/d\xi dt$ cross section measured by CDF at Tevatron energy, as given in \cite{GM}.
In particular, at $\sqrt s=1.8$ TeV, with a proton momentum fraction transferred through the pomeron of $\xi=1-x_L=0.01$  and $-t=0.05$ GeV$^2$, the CDF collaboration claim 
\be
d\sigma/d\ln\xi dt\simeq 2 ~{\rm mb/GeV^2},
\ee
 while TOTEM at $\sqrt s=7$ TeV  gives about\footnote{To obtain this estimate we have 
 divided the cross section (3.3 mb for single proton dissociation of {\em both} incoming protons) 
 measured in the central $8 <M_X < 350$ GeV interval by the size ($\Delta\ln 
 M^2_X=7.56$) of the rapidity interval, and accounted for the corresponding $t$-slope 
 ($B=8.5$ GeV$^{-2}$) observed by TOTEM~\cite{TO4}. Thus we obtain  $d\sigma_{\rm SD}/d\ln\xi 
 dt=(3.3~{\rm mb}/2/7.56)\times 8.5~$GeV$^{-2}\times\exp(-8.5\times 0.05)=1.2$ mb/GeV$^2$.} 
 1.2 mb/GeV$^2$, for the same mass of the diffractive state, $M_X\sim 100 - 200$ GeV. That is, TOTEM has a cross section about factor 1.7 smaller than CDF. On the other hand, naively, we would expect that the value of the diffractive dissociation cross section to increase with energy.

\item Next the cross section $d\sigma_{\rm SD}/d\ln\xi$ in the first (3.4 to 8 GeV) $M_X$ interval is more than twice larger than that in the central interval. (Indeed, dividing the TOTEM preliminary cross sections presented in Table 1 by the size of the $\ln M^2$ intervals (1.71 and 7.56) we find  $d\sigma_{\rm SD}/d\ln\xi=1.05$ mb and 0.44 mb for the first and the second mass intervals respectively.)
 Of course, according to the triple-Regge formula, a pomeron intercept $\alpha_P(0)>1$ leads to an increase of the cross section when $\xi$ decreases, but by the same argument we have to observe a larger cross section at the LHC than at the Tevatron, for the same value of $M_X$, contrary to the data. 

\item An analogous problem is observed for low-mass dissociation, where the cross section, $\sigma_{\rm SD}^{{\rm low}M_X}$, was about 30\% of the elastic cross section at CERN-ISR and fixed target energies \cite{Kaid}, whereas it turns out to be only 10\% at the LHC~\cite{TO2}. 
\end{itemize}

All these puzzles may be explained semi-quantitatively  by the fact that the values of the pomeron couplings are not fixed, but decrease with energy due to the growth of $k_t$ of the intermediate partons along the pomeron exchange ladder.

We attempt a simultaneous description of all these data within a two-channel eikonal framework, together with multi-pomeron interactions. We discuss elastic and low-mass dissociation in the next Section. Then in Sections \ref{sec:4} and \ref{sec:5} we discuss the description of data for high-mass single dissociation and double dissociation respectively. Although these discussions may seem to be self-contained analyses, we emphasize that they are just parts of a single `global' description.  We give a discussion in Section \ref{sec:6}, together with a summary of model predictions of high energy diffractive observables.

\section{Elastic scattering and low-mass dissociation}

These quasi elastic processes are described in terms of the Good-Walker formalism in which both of the incoming proton states are expressed as a linear sum over the diffractive eigenstates, $|p\rangle = \sum_i a_i |\phi_i\rangle$.

\subsection{Description of elastic scattering \label{sec:3}}
In terms of the G-W framework, the differential elastic cross section takes the form
\be
\frac{d\sigma_{\rm el}}{dt}~=~\frac{1}{4\pi}  \left| \int d^2b~e^{i\q_t \cdot \b} \sum_{i,k}|a_i|^2 |a_k|^2~(1-e^{-\Omega_{ik}(b)/2}) \right|^2,
\ee
where $-t=q_t^2$ and the opacity is driven by one-pomeron-exchange (between states $\phi_i$ and $\phi_k$ in the $b$-representation)
\be
\label{eq:ob}
\Omega_{ik}(s,b)=\int\frac{d^2q_t}{4\pi^2}\Omega_{ik}(s,q_t)e^{i\q_t \cdot \b}
\ee
with
\be
\Omega_{ik}(s,q_t)=g_i^N(t) g_k^N(t) \left(\frac s{s_0}\right)^{\alpha_P(t)-1}.
\label{eq:ot}
\ee
We use a two-channel eikonal; that is, two G-W diffractive eigenstates $i,k=1,2$. The normalization, Im$T=s\sigma$, is such that the pomeron-nucleon couplings 
\be
g_i^N=\gamma_i\sqrt{\sigma_0}F_i(t),
\label{eq:gamma}
\ee
where the form factors satisfy $F_i(0)=1$. Thus the cross section for the interaction of eigenstates $\phi_i$ and $\phi_k$, via one-pomeron-exchange, is
\be
\sigma_{ik}=\sigma_0 \gamma_i \gamma_k (s/s_0)^\Delta.
\ee
The form factors are parametrized as
\be
F_i(t)={\rm exp}((-b_i(c_i-t))^{d_i}+(b_ic_i)^{d_i}).
\label{eq:ff}
\ee
The $c_i$ term is added to avoid the singularity $t^{d_i}$ in the physical region of $t<4m^2_{\pi}$. Note that $F_i(0)=1$. 

The parameters $b_i,~c_i,~d_i$, together with the intercept and slope of the pomeron trajectory are tuned to describe the elastic scattering data, paying particular attention to the energy behaviour of 
low mass dissociation cross section. 
We first discuss the description of the elastic data. In order to correctly describe the dip region we must include the real part of the amplitude. We use a dispersion relation. For the even-signature pomeron-exchange amplitude this means
\be
A~\propto ~s^{\alpha} + (-s)^{\alpha} ~~~~~~~{\rm and~so~we ~have} ~~~~~~~\frac{{\rm Re}~A}{{\rm Im}~A}={\rm tan}(\pi\alpha /2),
\label{eq:RE}
\ee 
that is the usual signature factor. This formula is transformed into $b$-space, so that the complex opacities, $\Omega_{ik}(b)$ in (\ref{eq:ob}) can be constructed. For each value of $b$, that is for each partial wave $l$, we calculate $\alpha$ and determine Re$~A$ from (\ref{eq:RE}).

\begin{figure} 
\begin{center}
\vspace*{-6.0cm}
\includegraphics[height=21cm]{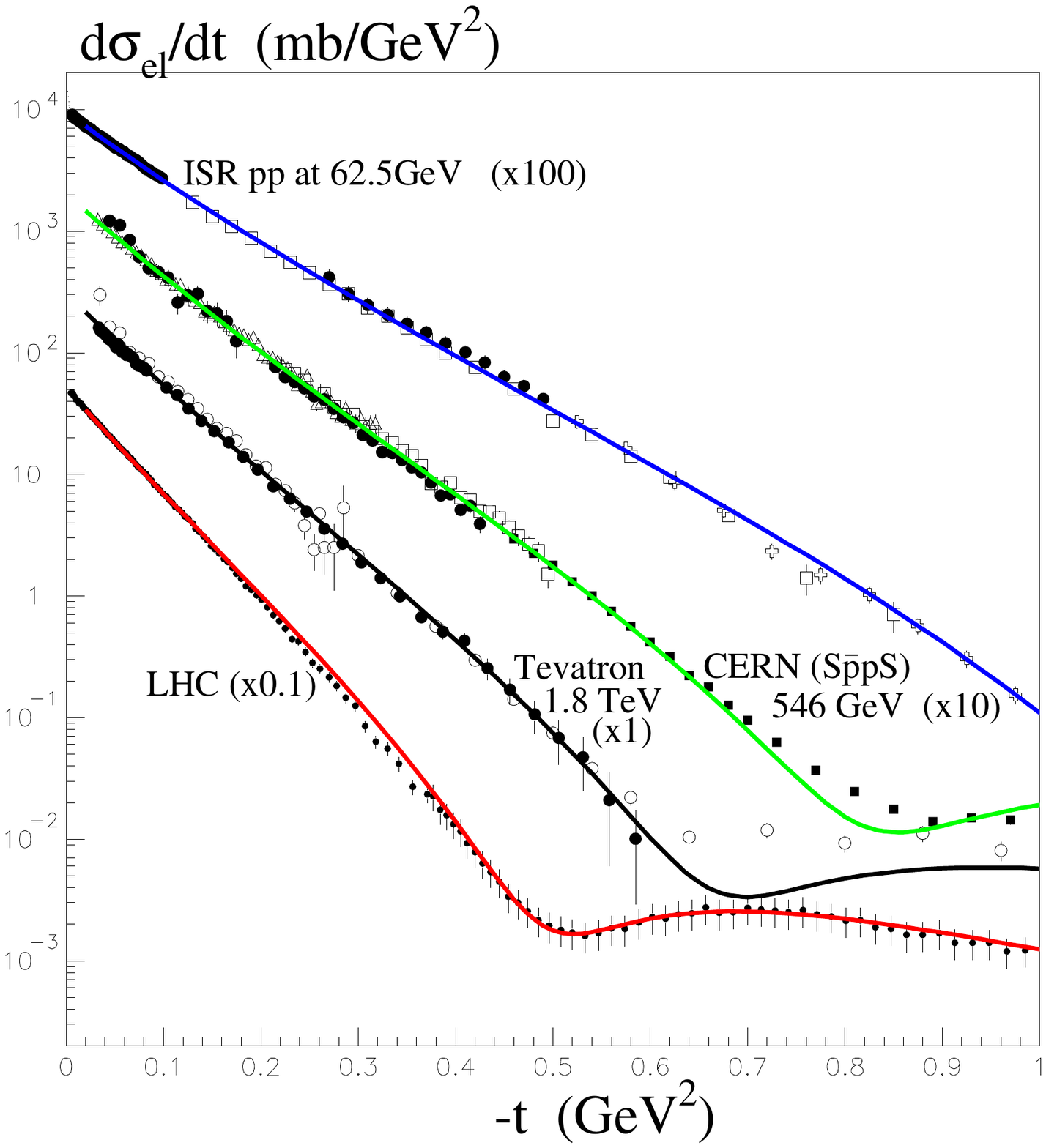}
\vspace*{-0.5cm}
\caption{\sf \sf The description of $pp$ or ($p{\bar p}$) elastic data. The references to the pre-LHC elastic data can be found in \cite{KMRLHC}. Here LHC refers to 7 TeV and the data are from \cite{TO1,TO1a}}
\label{fig:A}
\end{center}
\end{figure}

In order to reproduce the cross section in the diffractive dip region  we find that the form factors, (\ref{eq:ff}) have to have powers $d_1=0.52$ and $d_2=0.51$, close to the form used long ago by Orear {\it et al.}, $F={\rm exp}(-b\sqrt{t})$~\cite{Or}. The values of the other parameters are
$c_1=0.35,\, c_2=0.25,\, b_1=4.7,\, b_2=4.1$ in GeV units.
In addition we take $|a_1|^2=0.265$, with $|a_2|^2=1-|a_1|^2$, and $\sigma_0\equiv (g_N(0))^2=57$ mb, where $g_N(t)$ is the proton-pomeron coupling
  The resulting description of the elastic  data is shown in Fig. \ref{fig:A}.

The description of the proton-antiproton scattering at large $|t|>0.6$ GeV$^2$  is not perfect. This may be caused by the fact that we do not include secondary reggeon contributions. We also are not considering here a possible Odderon exchange contribution.
 
\subsection{Description of low-mass dissociation}
The next part of the `global' description that we discuss is low-mass dissociation.  Here the experimental information is a puzzle in that the cross section
$\sigma_{\rm D}^{{\rm low}M_X}$ goes from about $2-3$ mb at the CERN-ISR energy\footnote{The relevant experimental references are listed in \cite{KMRLHC}.} of 62.5 GeV to only $2.6\pm 2.2$ mb at 7 TeV at the LHC \cite{TO2}. Thus $\sigma_{\rm D}^{{\rm low}M_X}$ is about 30$\%$ of $\sigma_{\rm el}$ at 62.5 GeV and only about 10$\%$ at 7 TeV, whereas we would expect these percentages to be about the same for single pomeron exchange.
This problem was discussed in \cite{KMRLHC}, and its resolution involves more understanding of the decomposition of the G-W diffractive eigenstates $|\phi_i\rangle$.

First, recall some properties of the Good-Walker framework. If, for simplicity, we fix state $k$ and consider the dissociation of only one proton $|p\rangle = \sum_i a_i |\phi_i\rangle$, then
\be 
\sigma_{\rm el}=|\langle p|T|p\rangle|^2=\left( \sum_i |a_i |^2 ~ T_i\right)^2=\langle T\rangle ^2,
\label{eq:e}
\ee
where $\langle T \rangle$ denotes the average of $T_i$ over the probability distribution of the diffractive eigenstates.
On the other hand, if we include both the elastic process and proton dissociation, then
\be
\sigma_{\rm el}+\sigma_{\rm SD}=\sum_i | \langle \phi_i | T | p\rangle |^2=\sum_i|a_i|^2 ~T_i^2~=~\langle T^2 \rangle.
\label{eq:d}
\ee
That is, the cross section for dissociation,
\be
\sigma_{\rm SD}~=~ \langle T^2 \rangle -\langle T \rangle ^2 ,
\label{eq:dispersion}
\ee
is given by the dispersion of the scattering amplitude, $T$. If all the components of the initial proton are absorbed equally, then the diffracted superposition is equal to the initial one and the dissociation cross section is zero.

The generalisation to double dissociation is straightforward. For completeness we give the full expressions for the elastic and the `total' low-mass diffractive cross sections (analogous to (\ref{eq:e}) and (\ref{eq:d}) respectively)
\bea
\sigma_{\rm el}~=~  \int d^2b \left|~\sum_{i,k}|a_i|^2 |a_k|^2~(1-e^{-\Omega_{ik}(b)/2}) \right|^2,\\
\sigma_{\rm el+SD+DD}~=~  \int d^2b ~\sum_{i,k}|a_i|^2 |a_k|^2 ~\left|(1-e^{-\Omega_{ik}(b)/2}) \right|^2,
\eea
where SD includes the single dissociation of both protons.
So the low-mass diffractive dissociation cross section is
\be
\sigma_{\rm D}^{{\rm low}M}~=~\sigma_{\rm el+SD+DD}-\sigma_{\rm el}.
\ee

We are now ready to resolve the puzzle of the energy dependence of $\sigma_{\rm D}^{{\rm low}M_X}$. The pomeron-eigenstate $|\phi_i \rangle$ coupling
 is driven by the impact parameter separation, $\langle r\rangle$, between the partons in the $|\phi_i\rangle$ state. The well known example is so-called colour transparency, where the cross section $\sigma\propto \alpha_s^2\langle r^2\rangle$~\cite{CT1,CT2,CT3,CT4}. However, if the transverse size of the pomeron becomes much smaller than this separation, then the cross section (and coupling) will be controlled by the pomeron size, that is by the characteristic $k_t$ in the pomeron ladder. In this limit $\sigma\propto 1/k^2_t$. Therefore it is natural to choose the following parametrization for the pomeron-$|\phi_i\rangle$ couplings
\be
\gamma_i\propto \frac 1{k^2_P+k^2_i}, 
\label{eq:gam-i}
\ee 
where the $\gamma_i$ are defined in (\ref{eq:gamma}), with the normalization $(\gamma_1+\gamma_2)/2=1$. Here $k_P$ is the characteristic transverse momentum of the pomeron,which we expect to behave as
\be 
k^2_P= k^2_{P0}\left(\frac{s x^2_0}{s_0}\right)^D\ .
\label{eq:D}
\ee 
%({\bf Is $s_0$=1 GeV2?  What is $10^2$ doing here? Seems more natural to insert $10^4$?})\\
In other words, during the evolution in $\ln(1/x)$, due to the BFKL 
diffusion in $\ln k^2_t$ \cite{Lipatov}, the square of the characteristic momentum $k^2_P$ grows as a power $D$ of $1/x$. Of course, we do not expect that the whole available $\ln(1/x)$ (rapidity) space will be subject to  diffusion. Rather, we assume, that as $x$ decreases, the diffusion starts from some relatively low $x=x_0$ parton with $x_0=0.1$. That is, the rapidity space available for the $\ln k^2_t$ diffusion is not  $\ln(s/s_0)$, but is diminished by $\ln(1/x_0)$ from both sides. (As usual we use  $s_0=1$ GeV$^2$.) The typical transverse momentum of this (starting) parton, inside the state $\phi_i$, is denoted by $k_i$ in (\ref{eq:gam-i}). In our `global' model description we take $D=0.28$. The value of $D$ is related to the $s^\Delta$ behaviour, with $\Delta=0.2-0.3$, of resummed BFKL, which is mentioned in Section \ref{sec:pom} below. However, the relation is not direct. Rather, it is some approximation of the resummed BFKL diffusion in $\ln k_t$. For this reason we keep $D$ as a free parameter.

The parametrisation of $\gamma_i$ in (\ref{eq:gam-i}) is such that at very large energies all the $\gamma_i$ tend to the same value, so the dispersion shown in (\ref{eq:dispersion}) decreases leading to a smaller probability of low-mass proton dissociation, while at lower energies we tend to the naive expectation $\gamma_i \propto 1/k^2_i$. Actually the value of the additional transverse momenta $k_P$ in (\ref{eq:gam-i}) turns out to be rather small in the fit to the data --  $k_P/k_1=0.35$ and $k_P/k_2=0.17 $ at $\sqrt s=1800$ GeV.  Nevertheless the dissociation is slowed sufficiently with increasing energy such that we achieve values of the cross section $\sigma_{\rm D}^{{\rm low}M_X}$ which are compatible with the data -- namely, we find the model gives 2.6 mb at $\sqrt{s}=62.5$ GeV, and 3.8 mb at $\sqrt{s}=7$ TeV.

\subsection{Parameters of the `effective' pomeron trajectory \label{sec:pom}}

In the present approach we do not account explicitly for enhanced absorptive effects, which would renormalize the pomeron trajectory. Instead, we deal with an effective renormalized pomeron. Therefore  it is not surprising that the value $\Delta=0.12$ found for the effective pomeron is {\it larger} than 0.08 (the value obtained when the amplitude was parametrized by one-pole-exchange without any multi-pomeron corrections
 \cite{DL}), but is {\it smaller} than the intercept, $\Delta\sim 0.2\ -\  0.3$,
 expected for the bare pomeron of the resummed NLL$(1/x)$ BFKL approach 
 \cite{resum1,resum2,resum3}.  Indeed,
in comparison with the simple model, we explicitly account for the non-enhanced eikonal absorption which suppresses the growth of the amplitude with energy.  Therefore to describe the same data we need a larger intercept ($\Delta=0.12$).  On the other hand, since we do not explicitly include the enhanced diagrams (which would also slow down the growth of the cross section in the eikonal approach)  we anticipate a smaller effective intercept than that given by resummed BFKL.  Similar arguments apply to the slope of the effective trajectory, leading to a value\footnote{Besides the constant slope, $\alpha'$, of the pomeron trajectory, we insert the $\pi$-loop contribution as proposed in \cite{AG}, and as implemented in \cite{KMR18}} ($\alpha'=0.05 ~\GeV^{-2}$)  intermediate between the BFKL prediction ($\alpha' \gapproxeq 0$) and the old one-pole parametrization \cite{DL} ($\alpha'=0.25~ \GeV^{-2}$).

\section{High-mass dissociation  \label{sec:4}}

\begin{figure} 
\begin{center}
\vspace*{-1.0cm}
\includegraphics[height=12cm]{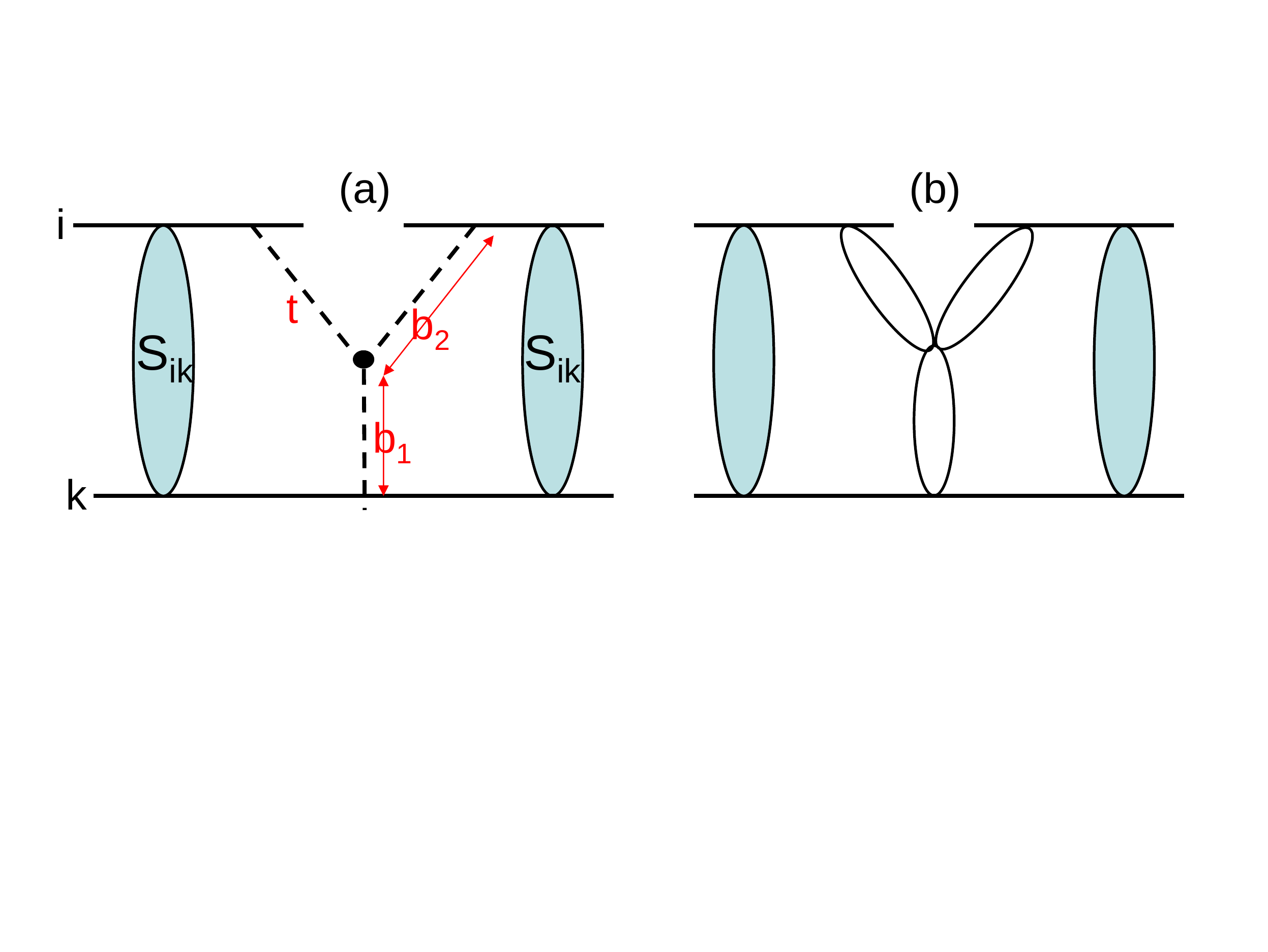}
\vspace*{-5.5cm}
\caption{\sf (a) A schematic diagram showing the notation of the impact parameters arising in the calculation of the screening corrections to the triple-pomeron contributions to the cross section; (b) a symbolic diagram of multi-pomeron effects. }
\label{fig:3Rb}
\end{center}
\end{figure}

The process $pp \to X+p$, where one proton dissociates into a system $X$ of 
{\it high-mass} $M$ is conventionally studied in terms of the triple-pomeron coupling, shown as the dot between the dashed lines  in Fig. \ref{fig:3Rb}(a). 
In the absence of absorptive corrections, the
corresponding cross section is given by
\be
\frac{M^2 d\sigma_{\rm SD}}{dtdM^2}~=~\frac{g_{3P}(t)g_N(0)g_N^2(t)}{16\pi^2}~\left(\frac{s}{M^2}\right)^{2\alpha(t)-2}~\left(\frac{M^2}{s_0}\right)^{\alpha(0)-1},
\label{eq:3P}
\ee
where $g_N(t)$ is the coupling of the pomeron to the proton and $g_{3P}(t)$ is the triple-pomeron coupling.
The value of the coupling $g_{3P}$ is obtained from a triple-Regge analysis of lower energy data.  Mainly they are the data on proton dissociation taken at the CERN-ISR with energies from $23.5 \to 62.5$ GeV.

The problem, with the above determination of $g_{3P}$, is that the value obtained is actually an effective vertex with coupling
\be
g_{\rm eff}~=~g_{3P}~\langle S^2\rangle
\ee
which already includes the suppression $S^2(b)=\exp(-\Omega(b))$ -- the probability that no other secondaries, simultaneously produced in the same $pp$ interaction, populate the rapidity gap region denoted by the + sign in $pp \to X+p$.  Recall that this survival factor $S^2$ depends on the energy of the collider.  Since the opacity $\Omega$ increases with energy, the number of multiple interactions, $N \propto \Omega$, grows\footnote{This is because at larger optical density $\Omega$ we have a larger probability of interactions.}, leading to a smaller $S^2$.  Thus, we have to expect that the naive triple-pomeron formula with the coupling \cite{abk,KKPT}, 
measured at relatively low collider energies will appreciably overestimate the cross section for high-mass dissociation at the LHC. A more precise analysis \cite{Luna} accounts for the survival effect $S^2_{\rm eik}$ caused by the eikonal rescattering of the fast `beam' and `target' partons.  In this way, a coupling $g_{3P}$ about a factor of 3 larger than $g_{\rm eff}$ is obtained, namely $g_{3P} \simeq 0.2g_N$, where $g_N$ is the coupling of the pomeron to the proton. The analysis of Ref. \cite{Luna} enables us to better allow for the energy dependence of $S^2_{\rm eik}$.

To account for the absorptive effect, it is easier to work in the impact parameter, $b$, representation. 
To do this we follow the procedure of Ref. \cite{Luna}. We first take Fourier transforms with respect to the impact parameters specified in Fig. \ref{fig:3Rb}(a). Then (\ref{eq:3P}) becomes 
\be
\frac{M^2 d\sigma_{ik}}{dtdM^2}~=~A\int\frac{d^2b_2}{2\pi}e^{i\vec{q}_t \cdot \vec{b}_2} \Omega_i(b_2)\int\frac{d^2b_3}{2\pi}e^{i\vec{q}_t \cdot \vec{b}_3} \Omega_i(b_3)\int\frac{d^2b_1}{2\pi} \Omega_k(b_1),
\label{eq:3Rb}
\ee
where $\Omega_i(b)$ is 
%described by 
the opacity corresponding to the interaction of eigenstate $\phi_i$ with a intermediate parton placed at the position of the triple-pomeron vertex, while $\Omega_k(b)$ describes the opacity of eigenstate $\phi_k$ from the proton which dissociates and interacts with the same intermediate parton. The normalization constant
\be
A=\pi^2/2g^2_N(0).
\ee
After integrating (\ref{eq:3Rb}) over $t$, the cross section becomes
\be
\frac{M^2 d\sigma_{ik}}{dM^2}~=~A\int\frac{d^2b_2}{\pi}\int\frac{d^2b_1}{2\pi} |\Omega_i(b_2)|^2 \Omega_k(b_1) \cdot S_{ik}^2(\vec{b}_2-\vec{b}_1),
\label{eq:result}
\ee
where here we have included the screening correction $S_{ik}^2$, which depends on the separation in impact parameter space, $(\vec{b}_2-\vec{b}_1)$, of states $\phi_i,\phi_k$ coming from the incoming protons
\be
S_{ik}^2(\vec{b}_2-\vec{b}_1)~\equiv~{\rm exp}(-\Omega_{ik}(\vec{b}_2-\vec{b}_1)).
\ee
If we now account for more complicated multi-pomeron vertices, coupling $m$ to $n$ pomerons, and assume an eikonal form of the vertex with coupling
\be
g^m_n=(g_N\lambda)^{m+n-2},
\label{eq:gmn}
\ee
then we have to replace $\Omega_i$ by the eikonal elastic amplitude and $\Omega_k$ by the inelastic interaction probability.  That is, instead of $\Omega_i(b_2)$ and $\Omega_k(b_1)$, we put
\be
\Omega_i \to 2(1-e^{-\Omega_i(b_2)/2}), ~~~~~~~~~~ \Omega_k \to (1-e^{-\Omega_k(b_1)}).
\label{eq:gmn2}
\ee
Fig. \ref{fig:3Rb}(b) symbolically indicates multi-pomeron couplings.
In (\ref{eq:gmn}), $g_N$ is the proton-pomeron coupling and $\lambda$ determines the strength of the triple-pomeron coupling.\footnote{In comparison with the (\ref{eq:ob},\ref{eq:ot}) expressions the formula for $\Omega_i$ contains an additional factor $\lambda/\pi$, that is we use (\ref{eq:ob}) with $\Omega_i(t)=g^N_i(t)g_{3P}(t)\exp(\Delta y_i(\alpha_P(t)-1))/\pi=g^N_i(t)\lambda g_N(0)\exp(B_{3P}t+\Delta y_i(\alpha_P(t)-1))/\pi$ where we assume the exponential dependence of $g_{3P}(t)\propto \exp(B_{3P}t)$ (for the each pomeron leg; see eqs.(4.6) and (4.7) of \cite{Luna}). Here $\Delta y_i$ is the rapidity interval between the proton ($i$) and the triple-pomeron vertex (intermediate parton); $\pi$ in the denominator comes from the definition of the
multi-Reggeon couplings; see an extra $\pi$ ($1/16\pi^2$ and not $1/16\pi$ as in usual elastic cross section) in (\ref{eq:3P}). $t$ dependence of the vertex is parametrized by conventional exponent with the slope $B_{3P}=0.7$ GeV$^{-2}$ for each pomeron leg which is in agreement with the last H1 data on diffractive $J/\psi$ production with the proton dissociation~\cite{H1} and with the results ($B_{3P}<1$/GeV$^2$ is small) of the previous triple-Regge analysis~\cite{KKPT,Luna}.}

\subsection{Implications of the TOTEM data for $\SD$ at high mass}
There are indications that the data for high-mass dissociation are not in agreement with the $M$ and $s$ dependence expected from the form of ${M^2 d\sigma}/{dtdM^2}$, based on (\ref{eq:3P}), assuming a {\it constant} $\lambda$.  From (\ref{eq:3P}) we see that the cross section 
should increase with decreasing 
 $M^2$ as 
 \be
 (1/M^2)^{2\alpha_P(t)-\alpha_P(0)-1}~\sim ~(M^2)^{-\Delta}.
 \label{eq:M}
 \ee
 However, the preliminary TOTEM data at $\sqrt s=7$ TeV,  give cross sections integrated over the $3.4 <M< 8$ and $8<M<350$ GeV mass intervals of 1.8 and 3.3 mb respectively \cite{TO4}.  This translates into a cross section ${M^2 d\sigma}/{dtdM^2}$  
 more than twice ($\sim2.4$) smaller for $M$ values in the second as compared to the first mass interval, whereas  (\ref{eq:M}) predicts only about a 60$\%$ increase.
 This observation indicates that the value of $\lambda$ (which specifies the multi-pomeron coupling) should be smaller in the second mass interval.  Secondly, since $\alpha_P(0)>1$, the cross section
for fixed $M^2$ should increase with energy ($\sqrt s$). On the other hand, the TOTEM result is about factor 1.7 less than that measured by CDF at the Tevatron ($\sqrt s=1.8$ TeV). Of course at the higher LHC energy we have a stronger suppression caused by the gap survival factor $S^2_{ik}$ (which was not included in the simplified expression (\ref{eq:3P})), but this is not enough to explain the discrepancy. (Note that the eikonal $S^2$ suppression is rather well fixed after the model was tuned to describe the elastic scattering and low-mass dissociation data.)

So we have phenomenological arguments in favour of introducing some energy dependence of $\lambda$, which specifies the multi-pomeron couplings via (\ref{eq:gmn}).  Since the $g^m_n$ coupling is a dimensionful quantity and the characteristic transverse momenta of the intermediate partons inside the pomeron ladder (i.e. the size of the pomeron) depend on the rapidity of corresponding partons, it looks natural to take 
\be
\lambda\propto 1/k^2_t(y)\ .
\label{eq:lambda1}
\ee
The diffusion in ln$k_t^2$ occurs from both the beam and target sides of the ladder. Following (\ref{eq:D}) we take $k_T^2 \propto (x_0/x)^D$ for diffusion from one side and $k_T^2 \propto (x_0/x')^D$ from the other side, where $xx's=\langle m^2_T\rangle$, which we take equal to $s_0=1~\GeV^2$.
So we parametrize $k_t(y)$ by
\be
k^2_t=k^2_0
 \left(\left(\frac{x_0}x\right)^D+\left(\frac{x_0}{x'}\right)^D\right)\ ,
\label{eq:lambda2}
\ee 
where we take the same $D=0.28$ and evolve from the same starting point $x_0=0.1$ as (\ref{eq:D}) for $\gamma_i$ of (\ref{eq:D}). We calculate  $x'$ as $x'=s_0/xs$ with $s_0=1$ GeV$^2$.  If $x>x_0$ we replace the $x_0/x$ ratio by 1, and similarly for $x'$.  

After we introduce the dependence of the multi-pomeron couplings on $x$, via 
 (\ref{eq:lambda1}) and (\ref{eq:lambda2}), the values obtained for the single proton 
 dissociation cross section (integrated over the three mass intervals used by TOTEM \cite{TO4}) are shown in Table \ref{tab:1}. 
\begin{table} [h]
\begin{center}
\begin{tabular}{|l|c|c|c|}\hline
 Mass interval (GeV) &   (3.4,~8) & (8,~350)&  (350, 1100)   \\ \hline
  Prelim. TOTEM data & 1.8 & 3.3  & 1.4   \\
 Present model & 2.3  & 4.0 & 1.4 \\
 \hline

\end{tabular}
\end{center}
\caption{\sf The values of the cross section (in mb) for single proton dissociation (integrated over the three    indicated mass intervals) as observed by TOTEM \cite{TO4}, compared with the values obtained in the present model. Recall that TOTEM claims that their preliminary measured cross sections have about 20\% error bars. }
\label{tab:1}
\end{table}
We see that the agreement with the mass dependence of the TOTEM data is now satisfactory. The $t$-slopes, defined by
\be
d\sigma_{\rm SD}/dt~\propto~e^{-B|t|},
\ee
evaluated, using the present model, for the interval $0.02<|t|<0.1~\GeV^2$, for the three mass TOTEM intervals are $B=8.5, ~ 7.2, ~ 6.0 ~\GeV^{-2}$ respectively (the preliminary TOTEM slopes are $B=10.1,~8.5,~6.8~\GeV^{-2}$; in agreement with the theoretical results within the experimental 15\% error bars.) .

To obtain the model predictions listed in Table \ref{tab:1}, we have included in the last mass interval the contribution of the secondary RRP term using the value of the RRP vertex found in the triple-Regge fit of \cite{Luna}. In the other two mass intervals such a contribution is negligible (less than 0.02 mb). We do not include the PPR contribution since it is dual to the low-mass proton excitations, which in our approach are accounted for in terms of the G-W diffractive eigenstates. 

In the present analysis we have taken $\lambda$ of (\ref{eq:gmn}) to be energy dependent, However, we find $\lambda=0.18$ at relatively low energies when both $x>x_0$ and $x'>x_0$ such that $\lambda$ ceases to be energy dependent. This value is in agreement with the previous triple-Regge analysis of \cite{KKPT,Luna}.

\subsection{Tension between high-mass single dissociation data}

Although TOTEM have made the most detailed observations of high-mass single proton dissociation in high energy $pp$ collisions, the present `global' diffractive  model has been tuned to simultaneously describe the TOTEM data {\it together} with earlier measurements of single dissociation. Here we compare with the description of measurements made by CDF at the Tevatron, and, later, in Section \ref{sec:5.3}, we show the description of information obtained by ATLAS \cite{atl}.

\begin{figure} 
\begin{center}
\vspace{-6.cm}
\includegraphics[height=14cm]{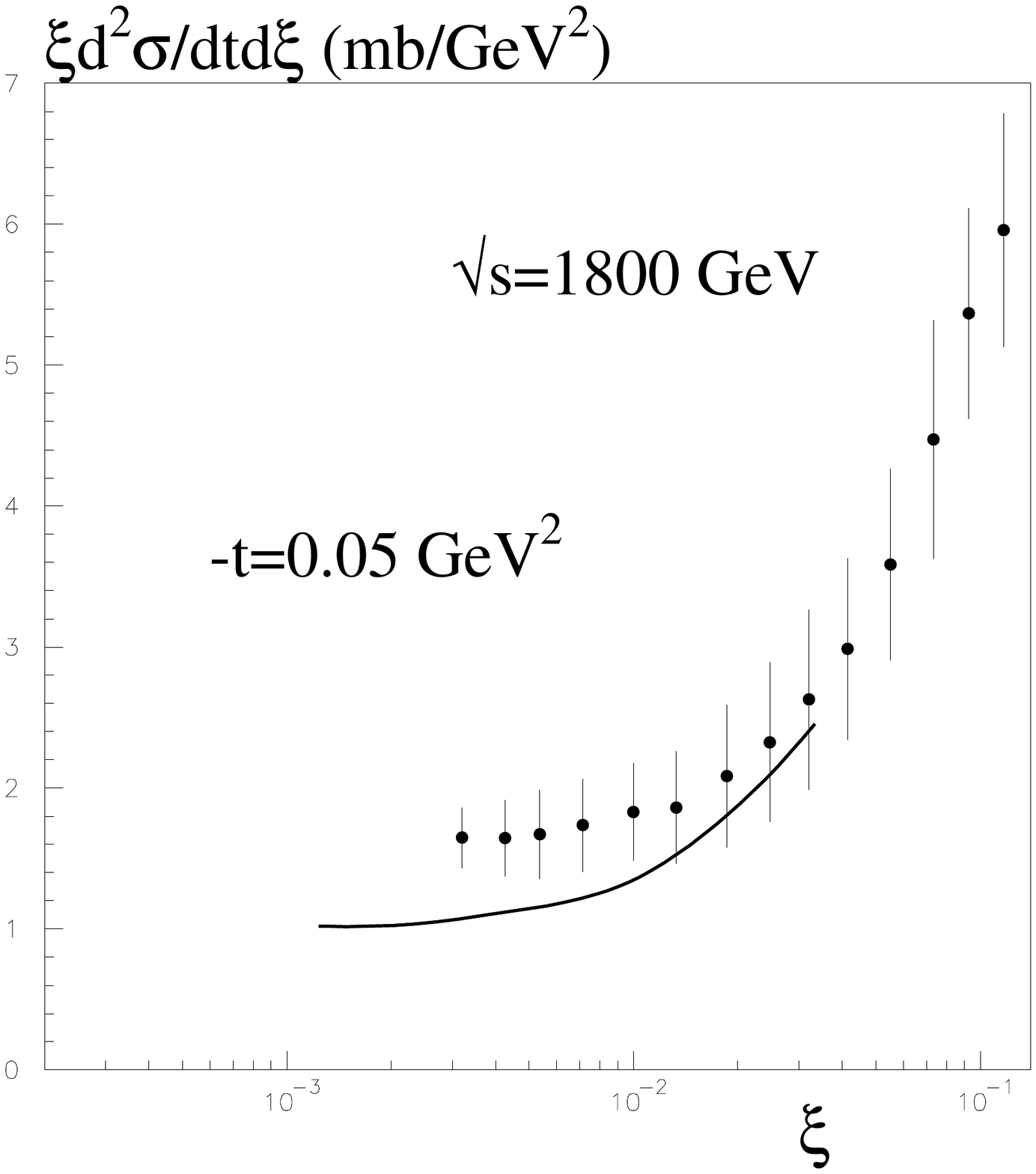}
\vspace*{-.5cm}
\caption{\sf The comparison of the model with data for single proton dissociation measured by the CDF collaboration, given in \cite{GM} but not including a normalisation uncertainty of about 10-15\%. The inclusion of the secondary Reggeon contribution RRP is responsible for the rise of the curve for $\xi$ increases.}
\label{fig:CDF}
\end{center}
\end{figure}

The comparison of the model with the cross section of single proton dissociation observed by the CDF collaboration at $\sqrt{s}=1800$ TeV and $-t=0.05 ~\GeV^2$ is shown in Fig. \ref{fig:CDF}.  We see that the agreement with the CDF data is not particularly good. However, note that: (a) there is some tension between the TOTEM data on the one hand, and CDF results (as well as those of ATLAS and CMS) on the other hand, which enforce us to tune the parameters in such a way that we overestimate the TOTEM single dissociation data, but simultaneously underestimate CDF, ATLAS and CMS cross sections, (b) actually these results were not published by the CDF collaboration, but were published in a separate paper by Goulianos-Montanha \cite{GM} and a normalization uncertainty of about 10 - 15\% were not included in the error bars.

\section{Factorisation and Double Dissociation \label{sec:5}}
 
 The recent TOTEM measurement of high energy double dissociation \cite{TO3} opens the way to study the relation between elastic, single dissociation and double dissociation cross sections.
\subsection{Naive factorisation}
\begin{figure} 
\begin{center}
\vspace{-3.cm}
\includegraphics[height=10cm]{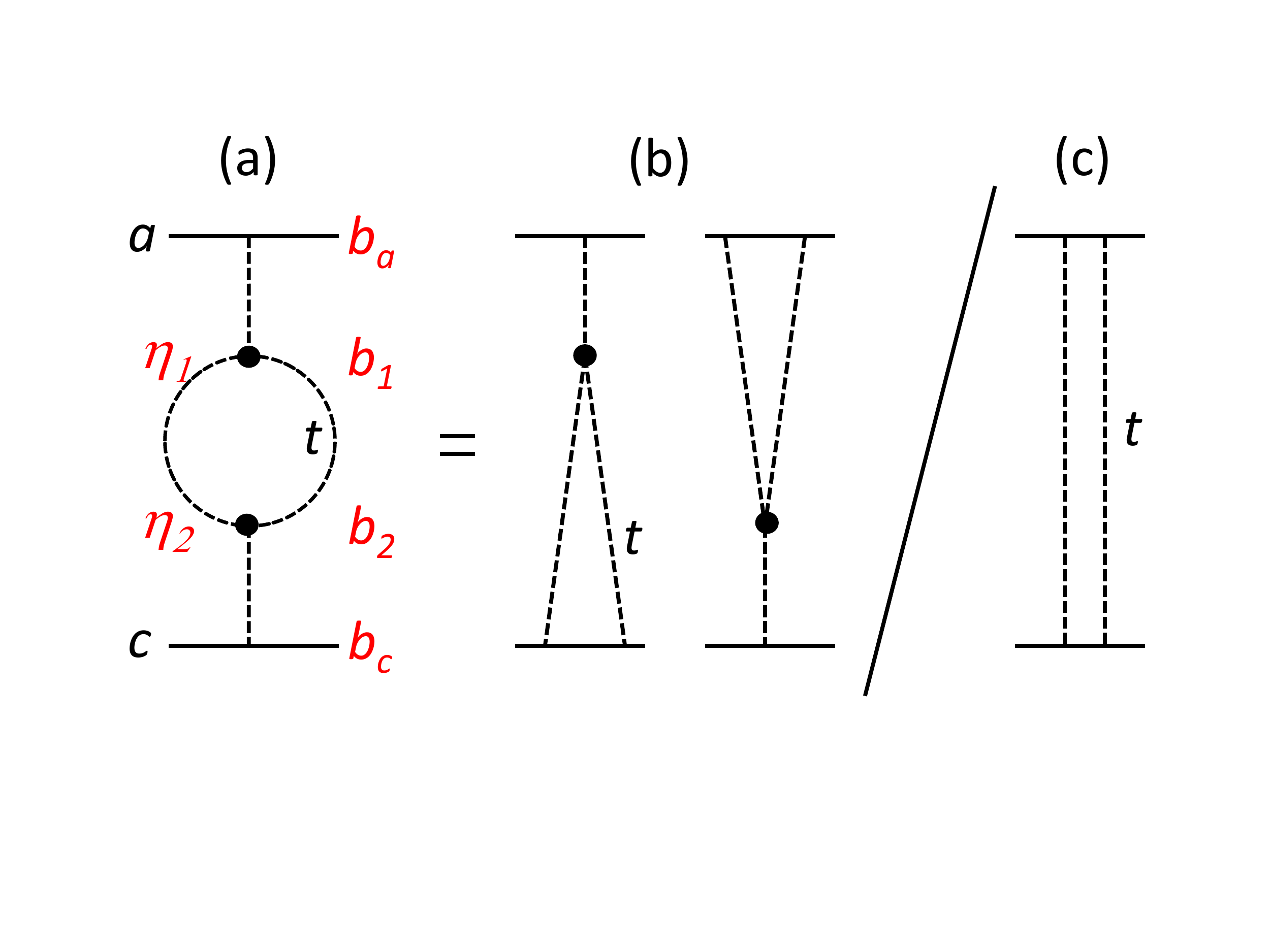}
\vspace*{-2.5cm}
\caption{\sf A pictorial representation of the naive factorization formulae of (\ref{eq:1}) and (\ref{eq:2}), resulting from the simplest pomeron exchange diagrams for (a) DD, (b) SD*SD and (c) elastic $ac$ scattering. It is convenient to evaluate the dissociation cross sections in impact parameter space, so we also show the variables $b_i$. }
\label{fig:3}
\end{center}
\end{figure}

 Within the framework of RFT, the simplest Reggeon diagram which describes the cross 
 section of high-mass diffractive double dissociation at high energies is the pomeron exchange diagram shown in 
Fig. \ref{fig:3}(a). As clear from Fig. \ref{fig:3}, it is natural to expect the 
 factorization relation
\be
\frac{d\DD}{dtd\eta_1d\eta_2}~~=~~\frac{d\SD}{dtd\eta_1}~\frac{d\SD}{dtd\eta_2}~/~\frac{d\el}{dt}.
\label{eq:2}
\ee
to be valid. Note that relation (\ref{eq:2}) is written for the differential cross section for some fixed value of the square of the momentum transfer $t$, and not for the cross sections integrated over $t$. The corresponding naive integrated factorisation relation is
\be
\DD~=~\frac{(\SD)^2}{\el}~~~~~~~~~~{\rm or}~~~~~~~~~~\frac{\DD~\el}{(\SD)^2}~=~1,
\label{eq:1}
\ee
where here $\SD$ is the single dissociation cross section from {\it one} proton, not the sum of both dissociations.
Before we compare the factorisation relation with the cross sections obtained by the  TOTEM collaboration  at $\sqrt{s}=7$ TeV, we must include some obvious violations expected for the naive form (\ref{eq:1}).

First, the relation is violated by the different $t$-slopes, $B$, of the elastic, single and double dissociation cross sections. Indeed, at 7 TeV the corresponding slopes are:
 $B_{\rm el}\simeq 20$ GeV$^{-2}$\cite{TO1,TO1a}, $B_{\rm SD}\simeq 10$ GeV$^{-2}$ for the lowest mass interval\footnote{To be specific, the preliminary values of the slopes observed by TOTEM \cite{TO4} in their three mass intervals are $B_{\rm SD}= 10.1,~8.5,~6.8~$ GeV$^{-2}$ respectively, with 15\% errors.} in \cite{TO4}, and the 
 estimated slope 
 \be
 B_{\rm DD}~\simeq ~2B_{3P}+2\alpha'_P|\eta_1-\eta_2|~=~ 2.4 
 ~\GeV^{-2}.
 \ee
 corresponding to the TOTEM experimental kinematics with 
 $|\eta_1-\eta_2|\sim 10$. For the estimate of $B_{\rm DD}$ we take the   
value  $\alpha'_P=0.05$ GeV$^{-2}$
obtained in Section \ref{sec:pom} to describe the elastic proton-proton cross section, 
 and we put the slope of the triple-pomeron vertex $ B_{3P}=0.7$ GeV$^{-2}$. Thus we already expect a violation of the naive relation (\ref{eq:1}) by a factor
 $B_{\rm SD}^2/B_{\rm el}B_{\rm DD}\sim 2$.

 More serious are the role of the 
 eikonal rapidity gap survival factors $S^2_{\rm eik}$. Both the single and the double 
 dissociative cross sections are suppressed by $S^2$. However, $S^2_{\rm SD}$ enters 
 (\ref{eq:1}) as the square of $S^2_{\rm SD}$, while $S^2_{\rm DD}$ enters as the first power. %Recall that 
The elastic 
scattering  cross section, which results from unitarity, has no explicit $S^2$ suppression, but, after accounting for the multi-pomeron diagrams, its value becomes less than that given by  single pomeron pole exchange. Using the `elastic' parameters of our model (given in Section \ref{sec:3}) we find a suppression of $d\sigma/dt|_{t=0}$ by a factor of about $ 6.8$. Moreover,   double dissociation occurs typically at somewhat larger values of the impact parameter, $b$, so $S^2_{\rm DD}>S^2_{\rm SD}$, see, for example, \cite{KMR18}.  These observations all lead to the left-hand-side being larger than the right-hand-side of (\ref{eq:1}). 

Thus, it is not a surprise to find sizeable breaking of naive factorisation. The question is whether we can account for the actual observed size of the breaking. Using the present model we find $S^2_{\rm SD}\simeq 0.08$ and a twice larger $S^2_{\rm DD}\simeq 0.16$. Thus, including the suppression of the elastic cross section and the slope factor, our estimate so far is
\be
\frac{\DD~\el}{(\SD)^2}~\simeq~\frac 2{6.8} ~\frac{0.16}{(0.08)^2}~\simeq 7.3.
\label{eq:rth}
\ee
 On the other hand, the TOTEM data give a much smaller violation of factorisation
 \be
\frac{\DD~\el}{(\SD)^2}~\simeq~\frac{0.116 \times 25}{(0.9)^2}~\simeq 3.6,
\label{eq:rexp}
\ee
 where here we use $\DD=0.116$ mb \cite{TO3},
 $\SD=1.8/2=0.9$ mb \footnote{The TOTEM result of 1.8 mb corresponds to single dissociation of {\em both} 
 protons, in the {\em same} rapidity interval as used for their measurement of $\DD$.}\cite{TO3} and $\el=25$ mb\cite{TO1,TO1a}.

Our model already gives satisfactory values for $\el$ and $\SD$.
Below, we therefore consider the possibility that a value of $\DD$ consistent with the TOTEM results can be obtained by the inclusion  of more detailed properties of our present model: the forms of the distributions in $b$-space, the multi-pomeron effects etc.  The multi-pomeron vertices were already included in our description of high-mass single dissociation, see eqs. (\ref{eq:gmn}) and (\ref{eq:gmn2}).

\subsection{Double dissociation and multi-pomeron contributions}
We start with the simplest expression for the double-dissociative cross section, corresponding to the process $pp \to X_1+X_2$ diagram shown in Fig. \ref{fig:3}(a); that is
\be
\frac{M^2_1 M^2_2 d\DD}{dtdM^2_1 dM^2_2}~=~\frac{g^2_{3P}(t)g_N^2(0)}{16\pi^3}~\left(
\frac{M^2_1}{s_0}\frac{M^2_2}{s_0}\right)^{\alpha_P(0)-1}
e^{2|\eta_2-\eta_1|(\alpha_P(t)-1)}\ ,
%~\left(\frac{M^2}{s_0}\right)^{\alpha(0)-1},
\label{eq:4DD}
\ee
where we have neglected the survival factor $S^2$. Here $M_1$ and $M_2$ are the masses of the dissociating systems from the two colliding protons, and the $\eta_i$ are the (pseudo)rapidities shown on the diagram. If we now integrate over the square of the momentum transferred, $t$, around the pomeron loop, and express
the opacities as a functions of their impact parameters, then (\ref{eq:4DD})
 takes the form
\be
\frac{d\DD}{d\eta_1d\eta_2}~=~\int dt\frac{d\DD}{d\eta_1d\eta_2 dt}~=~ \frac 1{g^4_N}\int d^2b_1 d^2b_2 d^2b_c
\Omega_{c2}(\Omega_{12}/2)^2 \Omega_{1a}e^{-\Omega_{ac}|\b_a-\b_c|}\ ,
\label{eq:Dop}
\ee
where now we have included the rapidity gap survival factor 
\be
S^2=\exp(-\Omega_{ac}(|\b_a-\b_c|)).
\ee
The notation $(a,~1,~2,~c)$ is specified in the diagram \ref{fig:3}(a). Here the opacities $\Omega_{1a}$ and $\Omega_{c2}$, between the nucleon and the corresponding triple-pomeron vertex, are defined as in (\ref{eq:ob}) and (\ref{eq:ot}), but since the vertex $g_{3P}=\lambda g_N $, the corresponding opacity contains an additional factor $\lambda$. In the same way, the opacity between the two triple-pomeron vertices contains a factor $(\lambda/\pi)^2$; see the footnote below eq.(\ref{eq:gmn2}). In particular, assuming a pure exponential $t$ dependence, this opacity takes the form
\be
\Omega_{12}(b_{12})=g^2_N\frac{\lambda^2}{\pi^2}~e^{2\Delta|\eta_1-\eta_2|}~
\left(\frac{e^{-b^2_{12}/4B_{12}}}{4\pi B_{12}}\right)\ ,
\label{op12}
\ee
where the slope
\be
B_{12}\equiv B_{DD}=2B_{3P}+2\alpha'_P|\eta_1-\eta_2|.
\ee
The factor $1/g_N^4$ in the denominator of (\ref{eq:Dop}) arises because in our normalization each opacity $\Omega\propto g^2_N$, while cross section (\ref{eq:4DD}) is proportional to $g^4_N$ only.  

To account for the {\em multi}-pomeron vertices, we have to replace $\Omega_{c2}$ and $\Omega_{1a}$ by the inelastic interaction probabilities $(1-\exp(-\Omega_{c2}))$ and $(1-\exp(-\Omega_{1a}))$, while the factor $\Omega_{12}/2$ is replaced by the probability of elastic parton ``$12$'' scattering, that is by $(1-\exp(-\Omega_{12}/2))^2$.
 Note that, after this eikonal unitarization, we now have no divergency in $\DD$ even in the case of a zero slope $B_{12}$; that is, even for $B_{3P}=0$ and $\alpha'_P=0$. Such a divergency which occurs in (\ref{eq:Dop}), due to the divergency of the $t$ integral and the corresponding divergency of $\Omega_{12}$ for $b_{12}=0$,
is now protected by  the parton ``12'' scattering amplitude, $1-\exp(-\Omega_{12}/2)$.

\begin{figure} [h]
\begin{center}
\vspace{-4.cm}
\includegraphics[height=16cm]{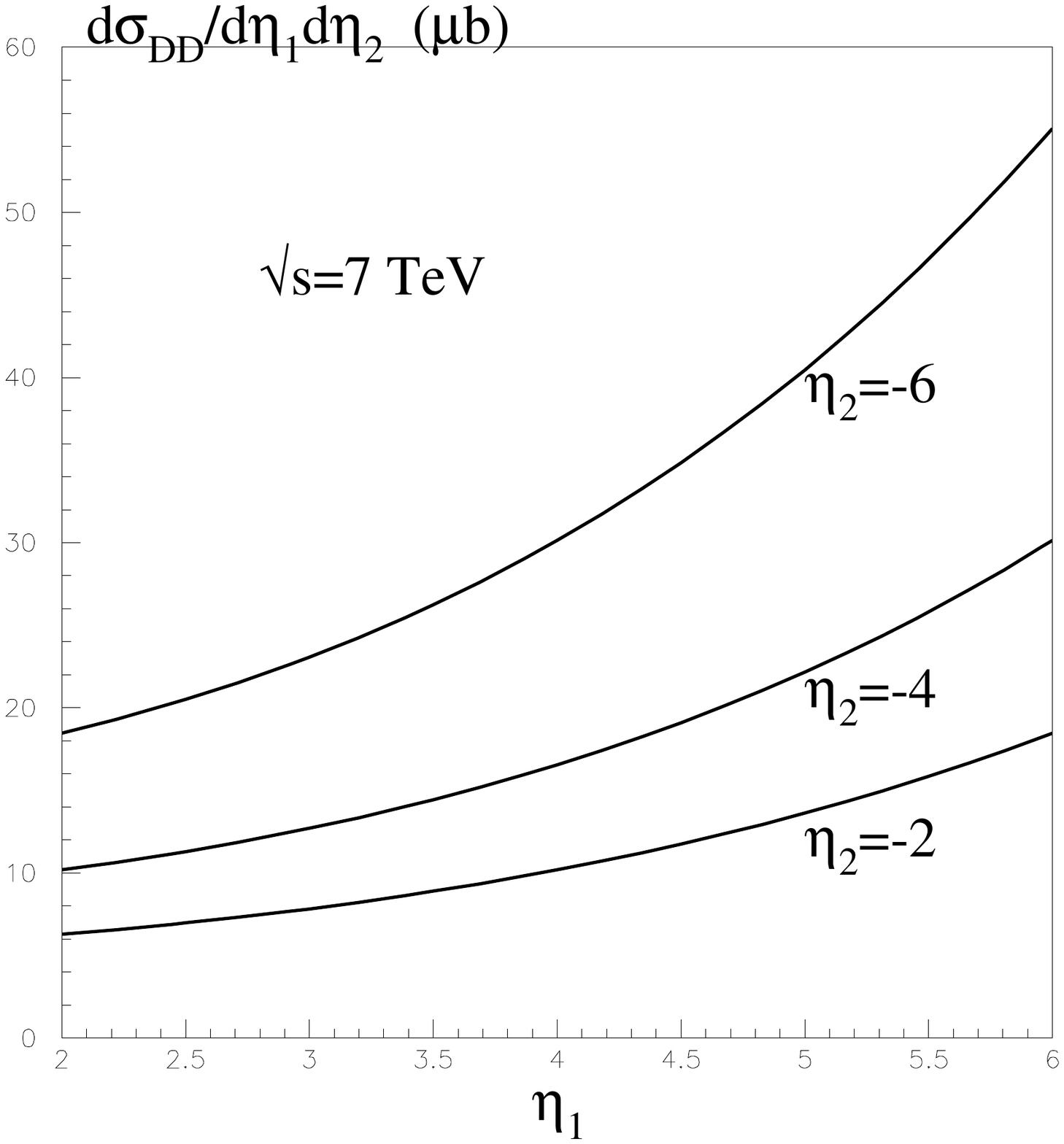}
\vspace*{-.5cm}
\caption{\sf The cross section (in $\mu$b) for double dissociation, $d\DD/d\eta_1 d\eta_2$, at the 7 TeV LHC, as a function of the position of the rapidity gap from $\eta_1$ to $\eta_2$, predicted by the present model which gives a `global' description of high energy elastic and diffractive data.}
\label{fig:C}
\end{center}
\end{figure}

In addition, the {\em multi}-pomeron vertices $g^m_n$ generate  gap survival factors with respect to ``1c'' and ``2a'' inelastic interactions. Overall this gives a screening factor
\be
 \exp(-\Omega_{2a}(|\b_2-\b_a|)-\Omega_{1c}(|\b_1-\b_c|))\ .
\label{s-enh}
\ee 
Thus, finally, we obtain
$$\frac{d\DD}{d\eta_1d\eta_2}~=~ \frac 1{g^4_N}\int d^2b_1 d^2b_2 d^2b_c
(1-e^{\Omega_{c2}})(1-e^{\Omega_{12}/2})^2 (1-e^{\Omega_{1a}})
~~\times$$
\be
% \frac 1{g^4_N}\int d^2b_1 d^2b_2 d^2b_c
%(1-e^{\Omega_{c2}})(1-e^{\Omega_{12}/2})^2 (1-e^{\Omega_{1a}})
~~~~~~~~~~\times~~ {\rm exp}(-\Omega_{ac}(|\b_a-\b_c|)-\Omega_{2a}(|\b_2-\b_a|)-\Omega_{1c}(|\b_1-\b_c|))\ . \label{eq:DDf}
\ee
Typical predictions for the differential cross section of double-dissociation, integrated over the $t$, are shown in Fig. \ref{fig:C}. They correspond to our `global' description of diffractive data, and account for the $k_t$ dependence of $\lambda$, keeping all the parameters determined as described in the previous Sections.

After the integration over the $-4.7~>\eta_2>-6.5$ and $4.7<\eta_1<6.5$ rapidity intervals covered by TOTEM, we obtain $\DD=145 \ \mu$b, close to the upper bound of the TOTEM measurement $116\pm 25\ \mu$b \cite{TO3}.   It is encouraging that the more physical and complicated structure of the present model largely reconcile the discrepancy between
(\ref{eq:rexp}) and (\ref{eq:rth}).

\subsection{Large Rapidity Gaps in central region, and SD and DD \label{sec:5.3}}

\begin{figure} [htb]
\begin{center}
\vspace*{-5.0cm}
\includegraphics[height=15cm]{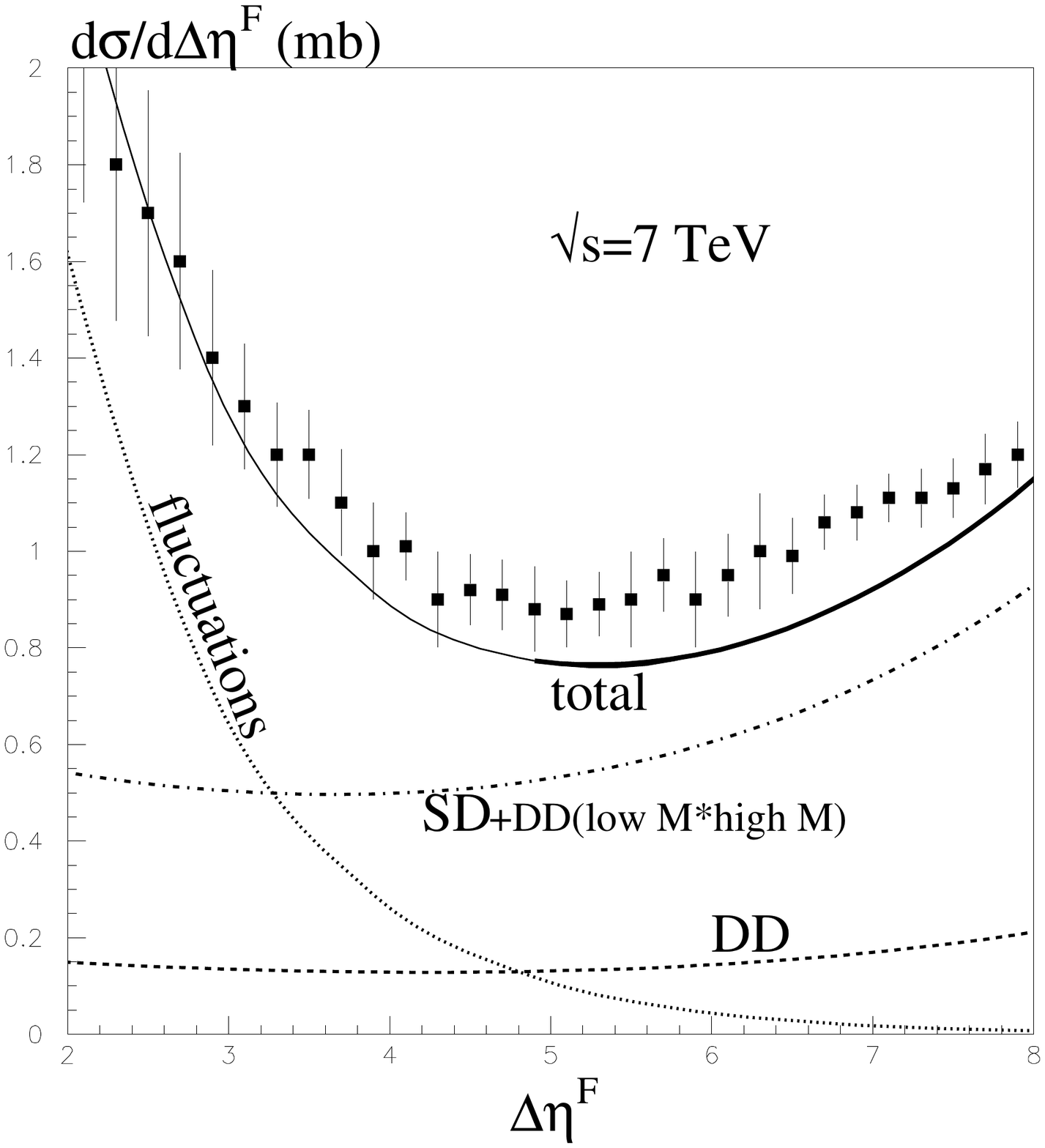}
\vspace{-0.0cm}
\caption{\sf The ATLAS \cite{atl} measurements of the inelastic cross section differential in rapidity gap size $\Delta\eta^F$ for particles with $p_T>200$ MeV. Events with small gap size ($\Delta\eta^F \lapproxeq 5$) may have a non-diffractive component which arises from fluctuations in the hadronization process \cite{FZ}. This component increases as $\Delta\eta^F$ decreases (or if a larger $p_T$ cut is used \cite{FZ,atl}). Therefore the curve at $\Delta\eta^F<5$ is shown by a thin line.  
The data with $\Delta\eta^F\gapproxeq 5$, which are dominantly of diffractive origin, are compared with the present `global' diffractive model.}
\label{fig:eta}
\end{center}
\end{figure}

The ATLAS \cite{atl} and CMS collaborations have measured the cross section of events with a large rapidity gap, $\Delta\eta^F$, which starts before the edge of the forward calorimeter
($\eta=4.9$ for ATLAS) and ends somewhere inside the opposite forward  calorimeter or in the tracking central detector. The ATLAS data are shown in Fig. \ref{fig:eta}, and correspond to measurements of the inelastic cross section differential in the size of the rapidity gap $\Delta\eta^F$ for particles with $p_T>200$ MeV. When $\Delta\eta^F$ decreases below about 5, the data are increasingly contaminated by fluctuations from the hadronisation process, but for $\Delta\eta^F \gapproxeq 5$ they are a measure of proton dissociation; in fact 
mainly of single proton dissociation. That is, the LRG actually starts just from a leading proton. However, we should not neglect the contribution of events where both protons dissociate, but the secondaries produced by one proton, say, the $M_X$-group, go into the beam pipe and are not seen in the calorimeter. In Fig. \ref{fig:eta} this double dissociation contribution is shown by the dashed curve.

It was demonstrated in \cite{FZ} that, depending on the particular mechanism of hadronization, the fluctuations may be able to account for the data at small $\Delta\eta^F$. To allow phenomenologically for such a possibility we  assume an exponential dependence
of this contribution, $\propto \exp(-a|\Delta\eta|)$ with $a=0.9$. If this term is normalized to the ATLAS data \cite{atl} then it gives the dotted line in Fig. \ref{fig:eta}. Recall however that the behaviour at  small $\Delta\eta^F$  
is strongly dependent on a hadronization model as discussed in \cite{FZ}.

\section{Discussion  \label{sec:6}}

The high energy diffractive data that are presently available cover a wide variety of processes. These include measurements of  the total and elastic $pp$ cross sections ($\sigma_{\rm tot},~ \el$), the elastic differential cross section $(d\el/dt$), 
the cross sections of low- and high-mass proton dissociation ($\SD^{{\rm low}M},~\SD^{{\rm high}M}$), the cross section of events where both protons dissociate $(\DD^{{\rm high}M}$), as well as the probability of inelastic events with a large rapidity gap ($d\sigma/d\Delta\eta$). 

Here, we demonstrate that all these diffractive data may simultaneously be described within the Regge Field Theoretic framework based on only one pomeron pole. However, to reach  agreement with the data, we have to include pomeron-pomeron interactions, arising from multi-pomeron vertices, and to allow for the $k_t(y)$ dependence of the multi-pomeron vertices. Recall that, due to the BFKL-type diffusion in $\ln k^2_t$ space,  together with the stronger absorption of low $k_t$ partons, the typical transverse momentum, $k_t$, increases with energy depending on the rapidity
 position of the intermediate parton or the multi-pomeron vertex. This $k_t(y)$ effect enables the model to achieve a relatively low probability of low-mass 
 dissociation of an incoming proton and to reduce the cross section of high-mass  
 dissociation in the central rapidity region in comparison with that observed closer 
 to the edge of available rapidity space -- both of which are features demanded by the recent TOTEM data.

Even though including the $k_t(y)$ dependence considerably improves the description of the dissociation data, the overall agreement with these data is not particularly good~\footnote{The imperfect description of the elastic proton-antiproton cross sections at larger $|t|$ may be due to the fact that at present we neglect the secondary reggeon contributions.}. This is mainly due to a {\em tension} between the TOTEM and the ATLAS, CMS, CDF results\footnote{Such a tension was also emphasized by S. Ostapchenko \cite{Ost}.}. It is not hard to improve the description of the TOTEM data on proton dissociation. We simply need a reduction of about 10 - 15\% of the starting value of $\lambda$, the parameter which specifies the muti-pomeron coupling. However, if we do this, we will even further underestimate the $M^2d\sigma/dM^2$  cross section at the Tevatron, and also the probability to have a LRG in the central rapidity region observed by the ATLAS and CMS\footnote{Recall that the CMS \cite{CMSdiff} cross section of dissociation integrated over the 12 - 394 GeV $M_X$ interval (close to, but in terms of $\ln M_X$, a bit smaller than, the interval (8 - 350) GeV chosen by TOTEM) is noticeably larger (4.3 mb) than that (3.3 mb) found by TOTEM.} groups. 
Here, we have tuned the model to give a compromise solution somewhere  between the CDF (ATLAS/CMS) and the TOTEM results.

It is also possible to obtain a lower value of $\SD$  integrated over the central of the three mass intervals used by TOTEM (while keeping the same cross sections in the low and large $M_X$ intervals)  by choosing a larger value of  the parameter $D$. However, if we were to do this then we would find that the probability of low-mass dissociation, $\sigma_D^{{\rm low }M}$, is too small (due to the small $\langle T^2\rangle -\langle T\rangle ^2$ dispersion caused by $\gamma_{1,2}\to 1$). Moreover, the model would then give an even steeper $d\sigma/d\Delta\eta^F$ behaviour of the LRG cross sections with increasing $\Delta\eta^F$. The model already has $d\sigma/d\Delta\eta^F$ growing  faster than the ATLAS and CMS data.

Here, we have adjusted the parameters of the model to give a reasonable description of all aspects of the available diffractive data.  If, instead, we had performed a $\chi^2$ fit to the  data, then the few dissociation measurements of TOTEM (values of $\SD$ in three mass intervals with 20\% errors, and one value of $\DD$) would have carried little weight.
On the  other hand, all the TOTEM data are self-consistent between themselves. Moreover, these data  reveal a very reasonable tendency of the $d\SD/d\xi$ dependence, close to that predicted in the model \cite{KMRsoft3} where the $k_t$ distribution of the intermediate partons inside the pomeron ladder, and the role of the transverse size of the different QCD pomeron components, were accounted for  more precisely.
 Therefore, we have presented the results of this `compromised' description (and not made a $\chi^2$ 
 fit) in order not to discard the interesting new information coming from 
the recent TOTEM measurements\footnote{We do not include in the present description the secondary Reggeon PPR contribution which is partly `dual' to that arising from the G-W diffractive eigenstates. In general, it should be considered in  future `global' diffractive analyses, but at present it does not change the situation qualitatively. So we prefer not to introduce the extra parameters.}.

For completeness, we give in Table \ref{tab:2} the values of some of the diffractive observables obtained from the present `global' description of diffractive high energy data.  We include, in particular, the values at collider energies relevant to experiments at the LHC.
\begin{table} [h]
\begin{center}
\begin{tabular}{|r|r|c|c|c|c|c|l|c|c|}\hline
$\sqrt{s}$ & $\sigma_{\rm tot}$ &   $\el$   &  $B_{\rm el}(0)$ & $\SD^{{\rm low}M}$ &     $\DD^{{\rm low}M}$ & $\SD^{\Delta\eta_1}$ &  $\SD^{\Delta\eta_2}$ &  $\SD^{\Delta\eta_3}$ & $\DD^{\Delta\eta}$ \\ \hline
  (TeV)&  (mb)  &    (mb)  & ($\GeV^{-2}$) & (mb) & (mb) & (mb) &(mb)&(mb)&($\mu$b)\\ \hline
    1.8 &  77.0  &   17.4  &   16.8  &  3.4  &   0.2 &  &  &  &   \\
    7.0 &  98.7  &    24.9 &   19.7  &  3.6  &   0.2 & 2.3 & 4.0  & 1.4 & 145 \\
    8.0 &  101.3  &   25.8 &   20.1  &  3.6  &   0.2 & 2.2 & 3.9 & 1.4 & 139 \\
   13.0 &  111.1  &   29.5 &   21.4 &  3.5  &   0.2  & 2.1 & 3.8 & 1.3 & 118 \\  
   14.0 &  112.7  &   30.1 &   21.6 &  3.5  &   0.2  & 2.1 & 3.8 & 1.3 & 115 \\
  100.0 &  166.3  &   51.5 &   29.4 &  2.7  &   0.1  &  &  &  &  \\
\hline
\end{tabular}
\end{center}
\caption{\sf The predictions of the present model for some diffractive observables for high energy $pp$ collisions at $\sqrt{s}$ c.m. energy. $B_{\rm el}(0)$ is the slope of the elastic cross section at $t=0$. Here $\SD$ is the sum of the single dissociative cross section of both protons. The last four columns are the model predictions for the cross sections for high-mass dissociation in the rapidity intervals used by TOTEM at $\sqrt{s}$=7 TeV: that is, $\SD$ for the intervals $\Delta\eta_1 = (-6.5,-4.7)$, $\Delta\eta_2 = (-4.7,~4.7)$, $\Delta\eta_3 = (4.7,~6.5)$, and $\DD^{\Delta\eta}$ is the double dissociation cross section in the two rapidity intervals $4.7<|\eta|<6.5$. At $\sqrt{s}$=7 TeV, the three `SD' rapidity intervals correspond, respectively, to single proton dissociation in the mass intervals $\Delta M_1 = (3.4,8)$ GeV, $\Delta M_2 = (8,350)$ GeV, $\Delta M_3 = (0.35,1.1)$ TeV, see Table \ref{tab:1}.}
\label{tab:2}
\end{table}

Recall that the slow rise of $\SD^{{\rm low}M}$  from a model value of 2.6 mb at the CERN-ISR energy to the value 3.6 mb at the LHC energy of $\sqrt{s}$=7 TeV is due to the growth of the characteristic momentum of the pomeron, $k_t^2 \propto s^D$, see (\ref{eq:D}). We noted that this behaviour is in accord with the TOTEM measurement of low-mass dissociation \cite{TO2}. Also, as just mentioned above, the energy dependence of the characteristic $k_t$ of the pomeron, which translates into a rapidity dependence, $k_t(y)$, is  in accord with the preliminary TOTEM measurements of $\SD$ in the three different mass (or rapidity) intervals, see Table \ref{tab:1}. The decrease of the cross sections for dissociation at $\sqrt{s}$=100 TeV, seen in Table \ref{tab:2}, is because we are beginning to approach the true black disk limit, where the probability of dissociation tends to zero,  while the effective
$\alpha'_{\rm eff}=\frac 12 dB_{\rm el}/d\ln s$ of elastic slope  increases.  

The values listed in Table \ref{tab:2} for $\sqrt{s}$=7 TeV are highly constrained by the recent measurements at the LHC. These measurements therefore largely determine the high energy predictions of the model. When more precise and extensive diffractive data become available, and the tensions between data sets are reduced, the model predictions may have to be adjusted.

\section*{Acknowledgements}

We thank Kenneth Osterberg, Risto Orava, Paul Newman and Sergey Ostapchenko for discussions.  MGR thanks the IPPP at the University of Durham for hospitality. This work was supported by the grant RFBR 14-02-00004 
and by the Federal Program of the Russian State RSGSS-4801.2012.2.

\bibliographystyle{JHEP.bst}
\bibliography{diff3}

\end{document}